\documentstyle[preprint,aps]{revtex}

\preprint{UCTP108.98}

\begin{document}
\draft

\title{One-loop finite temperature effective action of QED 
in the worldline approach}

\author{Igor A.~Shovkovy\thanks{
              On leave of absence from Bogolyubov Institute 
              for Theoretical Physics, Kiev 252143, Ukraine}}
\address{Physics Department, University of Cincinnati, 
         Cincinnati, OH 45221-0011, USA}
\date{July 17, 1998}
\maketitle

%%%%%%%%%%%%%%%%%%%%%%%%%%%%%%%%%%%%%%%%%%%%%%%%%%%%%%%%%%%%%%%%%%%%
\vspace{1cm}

\hrule
\begin{tabular}{ll}
{\bf Corresponding author:} & Dr. I.A.~Shovkovy  \\
{\bf Postal address:} 
                 & Physics Department            \\
                 & University of Cincinnati      \\
                 & Cincinnati, OH 45221-0011, U.S.A.     \\
{\bf E-mail:}    &   igor@physics.uc.edu \\
{\bf Telephone:} &   (513) 556-0520      \\
{\bf Fax:}       &  (513) 556-3425       \\
\end{tabular}

\hrule

\newpage
%%%%%%%%%%%%%%%%%%%%%%%%%%%%%%%%%%%%%%%%%%%%%%%%%%%%%%%%%%%%%%%%%%%%

\begin{abstract}
The one-loop finite temperature effective potential of QED in an 
external electromagnetic field is obtained using the worldline 
method. The general structure of the temperature dependent part 
of the effective action in an arbitrary external inhomogeneous 
magnetic field is established. The two-derivative effective 
action of spinor and scalar QED in a static magnetic background 
at $T\neq 0$ is derived.
\end{abstract}  


% 11.10.Kk Field theories in dimensions other than four 
% 11.10.Wx Finite-temperature field theory
% 11.15.-q Gauge field theories
% 11.10.Ef Lagrangian and Hamiltonian approach
% 12.20.-m Quantum electrodynamics
% 12.20.Ds Specific calculations 

%%%%%%%%%%%%%%%%%%%%%%%%%%%%%%%%%%%%%%%%%%%%%%%%%%%%%%%%%%%%%%%%%%%%%%%%
The use of the worldline formalism in quantum field theory 
has a long history \cite{BerM,BDVH,BBC,Pol,FG} which 
presumably starts from the original work by Schwinger 
\cite{Sch} where the so-called proper time was first 
introduced. The latter along with the representation of 
spinor degrees of freedom \cite{BerM} are the most 
important constituents of the method. 

In nineties, the worldline method has undergone a new 
wave of increased interest and extensive development 
\cite{BernK,Stras,McK1,McK2,Schmidt,vanH,RSS,GSh,FHSS}. 
Mainly, this was caused by the work of Bern and Kosower 
\cite{BernK} who established a set of rules for deriving
the multi-gluon amplitudes in the Yang-Mills field theory 
starting from the Polyakov perturbative formulation of 
the (heterotic) string theory in the limit of infinite 
tension. Soon after that, it was shown that the same rules 
could also be obtained from a perturbative evaluation 
of a particular worldline path integral \cite{Stras}.  
It was realized already in \cite{BernK} that, in many 
instances, the worldline approach has great advantages 
over the standard Feynman technique \cite{GSh,FHSS,BDK,AdlS}. 

It seems to be less known that the worldline method 
also allows a very simple and elegant generalization to 
finite temperatures \cite{McK1,McK3} (for a non-relativistic 
case see also \cite{GLSh}). In this letter, we apply the 
variant of the method of \cite{McK1} to the one-loop free
energy in an external electromagnetic field. The result 
for a constant external electromagnetic field is 
well known \cite{Dttr,EPS,CD}. Below we show how the 
same answer is obtained in the worldline approach. Besides 
the case of the constant field background, we calculate 
the derivative expansion of the thermal effective action 
of QED in a slightly inhomogeneous static magnetic field. 

%%%%%%%%%%%%%%%%%%%%%%%%%%%%%%%%%%%%%%%%%%%%%%%%%%%%%%%%%%%%%%%%%%%%%%%%
As is known, the one-loop free energy in QED 
reduces to computing the fermion determinant
\begin{eqnarray}
F^{(1)}(A)&=&-T \ln {\rm Det}(i\gamma^{\mu}{\cal D}_{\mu}-m)=
-\frac{T}{2}\ln{\rm Det}\left({\cal D}^{\mu}{\cal D}_{\mu}
+\frac{e}{2}\sigma^{\mu\nu}F_{\mu\nu}+m^2\right)=\nonumber\\
&=&-\frac{T}{2}\int\limits_{0}^{-i\beta} dx_{0}
\int d^{3}x\langle x|tr\ln\left({\cal D}^{\mu}{\cal D}_{\mu}
+\frac{e}{2}\sigma^{\mu\nu}F_{\mu\nu}+m^2\right) |x\rangle.
\label{ActGen2}
\end{eqnarray}
Here, by definition, ${\cal D}_{\mu}= \partial_{\mu}+ieA_{\mu}$ 
($\mu=0,\dots,3$) is the covariant derivative,
$\sigma^{\mu\nu}=i/2[\gamma^{\mu},\gamma^{\nu}]$, 
and $tr$ refers to the spinor indices of the Dirac 
$\gamma$-matrices. The states $|x\rangle$ are the 
eigenstates of a self-conjugate coordinate operator 
$x^{\mu}$. In our calculation we use the Minkowski metric.

By introducing the proper-time in a standard way \cite{Sch}, 
the free energy per unit volume is represented as follows,
\begin{equation}
\frac{F^{(1)}(A)}{V_3}=-\frac{i}{2} \int \limits^{\infty }_{0} 
\frac{d\tau}{\tau} e^{-im^2\tau} tr \langle x| \exp(-i\tau H) 
|x\rangle,
\label{free-e}
\end{equation}
where the second order differential operator $H$ is given by
\begin{equation}
H= {\cal D}^{\mu} {\cal D}_{\mu} + \frac{e}{2} \sigma^{\mu\nu} 
F_{\mu\nu}(x).
\label{operH}
\end{equation}
The matrix element $tr\langle z|\exp(-i\tau H)|y\rangle$ 
(which has the interpretation of the evolution operator of
a spinning particle), entering the right hand side of 
Eq.~(\ref{free-e}), allows a quantum mechanical path integral 
representation (compare with zero temperature case \cite{GSh})
\begin{equation}
tr \langle z|\exp(-i\tau H) |y\rangle = N^{-1}\int {\cal D} 
[x(t), \psi(t)] \exp\left\{ i\int \limits^{\tau}_{0} dt 
\left[ L_{bos}(x(t)) + L_{fer}(\psi(t),x(t)) \right] \right\},
\label{evol0}
\end{equation}
where $N$ is a normalization factor, and
\begin{equation}
L_{bos}(x)= -\frac{1}{4}\frac{dx_{\nu }}{dt}
\frac{dx^{\nu }}{dt}-eA_{\nu }(x) \frac{dx^{\nu }}{dt},
\label{L_bos}
\end{equation}
\begin{equation}
L_{fer}(\psi ,x)=\frac{i}{2}\psi_{\nu }\frac{d\psi^{\nu }}{dt}
-ie \psi^{\nu }\psi^{\lambda } F_{\nu \lambda }(x).
\label{L_fer}
\end{equation}
The integration in Eq.~(\ref{evol0}) goes over trajectories 
$x^{\mu}(t)$ and $\psi^\mu (t)$ parameterized by $t\in
[0,\tau]$. In addition, the definition of the integration 
measure assumes the following boundary conditions
\begin{equation}
x_{\nu }(0)=y_{\nu },\qquad 
x_{0}(\tau)=z_{0}~\mbox{mod}(i\beta),\qquad
x_{i}(\tau)=z_{i} \quad (i=1,2,3),
\qquad  \psi (0)=-\psi (\tau).
\end{equation}

Since the path integral includes the integrations over 
worldline trajectories with arbitrary integer windings
around the compact (imaginary) $x_0$-direction, the expression 
in Eq.~(\ref{evol0}) splits into the sum of path integrals 
labeled by the winding numbers. In case of spinor QED,
the weight factors of these separate contributions are given 
by $(-1)^n$ \cite{McK1}. Therefore,
\begin{eqnarray}
&& tr\langle z|\exp(-i\tau H) |y\rangle =
N^{-1}\sum_{n=-\infty}^{\infty}
(-1)^n \int {\cal D} [x^{(n)}(t) , \psi(t) ] 
\nonumber\\
&\times&\exp \left \{i\int \limits^{\tau}_{0} dt 
\left[L_{bos} \left(x^{(n)}(t) \right)
+L_{fer} \left( \psi(t), x^{(n)}(t) \right) \right] \right\},
\label{evol1}
\end{eqnarray}
where the boundary conditions $x^{(n)}_{\nu }(0)=y_{\nu }$ and 
$x^{(n)}_{\nu }(\tau)=z_{\nu }+in\beta\eta_{\nu 0}$ are assumed. 
Note, that there exists a similar representation for scalar QED 
as well. In contrast to the case at hand, in scalar QED, the 
integration over the Grassman field $\psi(t)$ is absent and all 
the weight factors are equal to 1 \cite{McK1}. 

%%%%%%%%%%%%%%%%%%%%%%%%%%%%%%%%%%%%%%%%%%%%%%%%%%%%%%%%%%%%%%%%%%%%%%%%
Let us first consider the finite temperature 
effective action of QED in a constant 
electromagnetic field. In this particular case, it is 
convenient to choose the vector potential in the 
following simple form,
\begin{eqnarray}
A_{\nu }(x) &=& \frac{1}{2} (x-y)^{\lambda }
F_{\lambda \nu }(y).
\label{Anu}
\end{eqnarray}
Then, the path integral in Eq.~(\ref{evol1}) becomes Gaussian,
and, therefore, the calculation can be done exactly. After 
performing the integration, we arrive at the result for the
diagonal matrix element of the evolution operator,
\begin{eqnarray}
tr \langle y|\exp(-i\tau H) |y\rangle &=&
-\frac{i} {4\pi ^{2}\tau^{2}} 
(e\tau K_{-}) (e\tau K_{+})\cot(e\tau K_{-}) 
\coth(e\tau K_{+})\nonumber\\ 
&\times& \sum_{n=-\infty}^{\infty} (-1)^n
\exp\left( i\frac{n^2\beta^2}{4}
\left(eF\coth(eF\tau)\right)_{00}
\right).
\label{evol2}
\end{eqnarray}
To disentangle the Lorentz indices in the exponent of the last 
expression it is convenient to make use of the matrices 
$A^{\mu}_{(j)\nu}$ that were originally introduced in \cite{BSh} 
and recently used in \cite{GSh}. In such a way, we arrive at the 
following identity,
\begin{eqnarray}
\left(eF\coth(eF\tau)\right)_{00}=
\frac{ (K_{+}^{2}-\vec{E}^2)eK_{-}\cot(e\tau K_{-})
+(K_{-}^{2}+\vec{E}^2)eK_{+}\coth(e\tau K_{+}) }
{K_{+}^{2}+K_{-}^{2}},
\label{iden}
\end{eqnarray}
where 
\begin{eqnarray}
K_{+}=\sqrt{\sqrt{ {\cal F}^2+{\cal G}^2 }+{\cal F} },
\qquad
K_{-}=\sqrt{\sqrt{ {\cal F}^2+{\cal G}^2 }-{\cal F} },
\end{eqnarray}
and we introduced the standard two invariants built of 
the electromagnetic field strength, namely, 
\begin{eqnarray}
{\cal F}=-\frac{1}{4} F^{\mu \nu }F_{\mu \nu },\label{F}
\qquad
{\cal G}=\frac{1}{8} \epsilon ^{\mu \nu \lambda \kappa } 
F_{\lambda \kappa } F_{\mu \nu } .
\end{eqnarray}
Notice that due to the presence of the heat bath, the Lorentz 
symmetry is explicitly broken. Therefore, obtaining of the 
electric field magnitude in the right hand side of Eq.~(\ref{evol2}) 
apart from the invariants is hardly unexpected.

After substituting Eq.~(\ref{iden}) into Eq.~(\ref{evol2}), we 
arrive at a convenient enough representation for the diagonal 
matrix element of the evolution operator. The latter, due to 
Eq.~(\ref{free-e}), gives the general result for the thermal 
effective action of QED in a constant electromagnetic field.

In the particular case of a pure magnetic background, the 
expression for the free energy per unit volume reads
\begin{eqnarray}
\frac{F^{(1)}(B)}{V_3}&=&-\frac{1}{8\pi ^{2}} \int 
\limits^{\infty }_{0} 
\frac{d\tau}{\tau^3} e^{-im^2\tau} 
(e\tau B) \cot(e\tau B)  
\sum_{n=-\infty}^{\infty} (-1)^n
\exp\left( i\frac{n^2\beta^2}{4\tau}
\right)
\nonumber\\
&=&\frac{|eB|^2}{8\pi ^{2}} \int 
\limits^{\infty }_{0} 
\frac{d\omega}{\omega^2} e^{-m^2\omega/|eB|} 
\coth(\omega) \sum_{n=-\infty}^{\infty} (-1)^n
\exp\left( -\frac{n^2\beta^2|eB|}{4\omega}
\right).
\label{free-mag}
\end{eqnarray}
This is exactly the result that was obtained in \cite{Dttr} and 
generalized in \cite{EPS,CD} for a non-zero chemical potential
(the case of a non-zero chemical potential was also considered 
in a non-relativistic case using a slightly different version of
the worldline method in \cite{GLSh}). 

In a pure electric background, on the other hand, the result 
takes a somewhat different form,
\begin{eqnarray}
\frac{F^{(1)}(E)}{V_3}&=&-\frac{1}{8\pi ^{2}} 
\int \limits^{\infty }_{0} 
\frac{d\tau}{\tau^3} e^{-im^2\tau}
(e\tau E)\coth(e\tau E) 
\sum_{n=-\infty}^{\infty} (-1)^n
\exp\left( i\frac{n^2\beta^2}{4}
(eE)\coth(eE\tau) \right)
\nonumber\\
&=&-\frac{|eE|^2}{8\pi ^{2}}
\int \limits^{\infty }_{0} 
\frac{d\omega}{\omega^2} e^{-im^2\omega/|eE|} 
\coth(\omega) 
\sum_{n=-\infty}^{\infty} (-1)^n
\exp\left( i\frac{n^2\beta^2|eE|}{4}
\coth(\omega) \right).
\label{free-elec}
\end{eqnarray}
We note that this is a complex expression. Usually, however, 
one expects for the free energy to be a real quantity under 
any circumstances. So, does this mean that the  result in 
Eq.~(\ref{free-elec}) is, in fact, wrong? The answer is no, 
of course. The point is that a system in an external electric 
field is not a truly equilibrium static system. Indeed, because
of the particle-antiparticle pair creation and the electric 
current induced by such a process, the system is out of 
equilibrium. 

%%%%%%%%%%%%%%%%%%%%%%%%%%%%%%%%%%%%%%%%%%%%%%%%%%%%%%%%%%%%%%%%%%%%%%%%
Now, let us consider the effective action in a more interesting 
case of a slightly inhomogeneous static magnetic field. Following 
the method developed in \cite{GSh}, we choose a version of the 
Fock-Schwinger gauge for the vector potential $A_{\mu}(x)$ as
follows,
\begin{equation}
A_{0}(x)=0, \qquad (x_{i}-y_{i})A_{i}(x)=0.
\end{equation}
The latter leads to the series
\begin{eqnarray}
A_{i}(x) &=& -\frac{1}{2} (x_{j} -y_{j}) 
F_{ji}(y) + \frac{1}{3} (x_{j}-y_{j})
(x_{l}-y_{l}) \partial _{l} F_{ji}(y)
\nonumber\\ &-& 
\frac{1}{8} (x_{j}-y_{j}) (x_{l}-y_{l}) 
(x_{k}-y_{k}) \partial _{l}\partial _{k} 
F_{ji}(y) + \ldots.
\label{eq:Ai}
\end{eqnarray}
With this choice of gauge, we arrite at a very convenient 
representation for the diagonal matrix element of the evolution 
operator, 
\begin{eqnarray} 
&&tr\langle y|U(\tau)|y \rangle=N^{-1}\sum_{n=-\infty}^{\infty}
(-1)^n \int {\cal D} [x^{(n)}(t) , \psi(t) ] \nonumber\\ 
&\times& \exp\left[i\int\limits^\tau_0 dt \left(
-\frac{1}{4}\frac{d x^{(n)}_{0}}{dt}\frac{d x^{(n)}_{0}}{dt}
+\frac{1}{4}\frac{d x^{(n)}_{i}}{dt}\frac{d x^{(n)}_{i}}{dt} 
-\frac{e}{2}x^{(n)}_{i} F_{ij}(y) \frac{d x^{(n)}_{j}}{dt}
+L_{bos}^{int}\left(x^{(n)}_{i}\right)\right)\right]
\nonumber\\ 
&\times&\exp\left[i\int\limits^{\tau}_{0}dt \left(
\frac{i}{2}\psi_{0} \frac{d\psi_{0}}{dt}
-\frac{i}{2}\psi_{i} \frac{d\psi_{i}}{dt}
-ie\psi_{i} \psi_{j} F_{ij}(y) 
+L_{fer}^{int}\left(x^{(n)}_{i},\psi_{i}\right) 
\right)\right],
\label{trU1}
\end{eqnarray}
where, as follows from Eqs.~(\ref{L_bos}), (\ref{L_fer}) and 
(\ref{eq:Ai}), the interacting terms, $L_{bos}^{int}(x)$ and 
$L_{fer}^{int}(x,\psi)$, containing spatial derivatives of 
$F_{ij}$, are given by
\begin{eqnarray}
L_{bos}^{int}(x)&=& 
- \frac{e}{3} F_{i j,k}\frac{d x_{i}}{dt} x_{j} x_{k}
+ \frac{e}{8} F_{i j,k l} \frac{d x_{i}}{dt} x_{j} x_{k} x_{l} 
+ \ldots,   
\label{eq:L2}\\
L_{fer}^{int}(x,\psi ) &=&
i eF_{i j,k} \psi_{i} \psi_{j} x_{k}
- \frac{ie}{2} F_{i j,k l} \psi_{i} \psi_{j} x_{k} x_{l} 
+ \ldots.
\label{eq:L3}
\end{eqnarray}

The integration variables in Eq.~(\ref{trU1}) are subject 
to the following boundary conditions, $x^{(n)}_{0,i}(0)=0$, 
$x^{(n)}_{0}(\tau)=in\beta$ and $x^{(n)}_{i}(\tau)=0$ (note 
that the fields $x^{(n)}_{0,i}(t)$ were preliminary shifted by 
$-y_{0,i}$).

A very useful property of the path integral in Eq.~(\ref{trU1}) 
is its factorization into two pieces. One of them contains only 
the time components of the fields and is, in fact, a Gaussian path 
integral. The other contains the interacting spatial components 
of the fields and, what is very important, does not depend on the 
winding number $n$. After performing the Gaussian integrations 
over $x^{(n)}_{0}(t)$ and $\psi_{0}(t)$, we obtain
\begin{eqnarray} 
&&tr\langle y|U(\tau)|y \rangle=
N^{-1}\sum_{n=-\infty}^{\infty}
(-1)^n\exp\left(i\frac{n^2\beta^2}{4\tau}\right)
\int {\cal D} [x_{i}(t), \psi_{i}(t) ] \nonumber\\ 
&\times& \exp\left[i\int\limits^\tau_0 dt \left(
\frac{1}{4}\frac{d x_{i}}{dt}\frac{d x_{i}}{dt} 
-\frac{e}{2}x_{i} F_{ij}(y) \frac{d x_{j}}{dt}
+L_{bos}^{int}\left(x_{i}\right)\right)\right]
\nonumber\\ 
&\times&\exp\left[i\int\limits^{\tau}_{0}dt \left(
-\frac{i}{2}\psi_{i} \frac{d\psi_{i}}{dt}
-ie\psi_{i} \psi_{j} F_{ij}(y) 
+L_{fer}^{int}\left(x_{i},\psi_{i}\right) 
\right)\right]. 
\label{factor}
\end{eqnarray}
This expression is, in fact, one of our most important results 
here. It states that the temperature dependent part of 
the evolution operator is exactly factorized from the part 
depending on an (arbitrary) inhomogeneous magnetic field. 
The latter, in its turn, puts a very strong restriction on the 
structure of the one-loop finite temperature effective action 
in QED. As is obvious, a similar restriction appears in scalar 
QED as well.

The further integration (over the spatial components of the fields) 
can be done only approximately. As is seen, this path integral 
is essentially the same as in zero temperature theory. So,
we could use the same perturbative expansion in the number of 
derivatives that was developed in \cite{GSh} for $T=0$. More than 
that, since the temperature dependent part of the effective action 
in Eq.~(\ref{factor}) is factorized exactly, we could use the known 
result \cite{GSh} without repeating all the tedious calculations.  

Therefore, assuming that the background magnetic field is directed
along the third spatial axis, we get the following expression 
for the two-derivative part of the finite temperature effective 
action,
\begin{eqnarray} 
\frac{F^{(1)}_{der}(B)}{V_3}&=&
\frac{e^2 \left(\partial_{\perp}B\right)^2}{(8\pi)^{2}|eB|} 
\int\limits_{0}^{\infty} \frac{d\omega}{\omega} 
e^{-m^2\omega/|eB|} 
\frac{d^3}{d\omega^3}\left( \omega \coth\omega \right)
\nonumber\\
&\times& \sum_{n=-\infty}^{\infty} (-1)^n
\exp\left( -\frac{n^2\beta^2|eB|}{4\omega}
\right),
\label{inB3+1}
\end{eqnarray} 
where $\left(\partial_{\perp}B\right)^2\equiv
\left(\partial_{1}B\right)^2+\left(\partial_{2}B\right)^2$.
The result in Eq.~(\ref{inB3+1}) differs from the analogous 
expression at $T=0$ (see Ref.~\cite{GSh}) only by the last 
factor containing the sum over the winding numbers. 
Again, as all the other results, this expression allows 
a straightforward generalization to 2+1 dimensional case as well 
as to scalar QED. For example, in 3+1 dimensional scalar QED,
the result reads 
\begin{eqnarray} 
\frac{F^{(1)scal}_{der}(B)}{V_3}&=&
-\frac{e^2 \left(\partial_{\perp}B\right)^2}{2(8\pi)^{2}|eB|} 
\int\limits_{0}^{\infty} \frac{d\omega}{\omega} 
e^{-m^2\omega/|eB|} 
\left(\frac{d^3}{d\omega^3}+\frac{d}{d\omega} \right)
\left( \frac{\omega }{\sinh\omega }\right)\nonumber\\
&\times&\sum_{n=-\infty}^{\infty}
\exp\left( -\frac{n^2\beta^2|eB|}{4\omega}
\right). 
\label{csB3+1}
\end{eqnarray}

%%%%%%%%%%%%%%%%%%%%%%%%%%%%%%%%%%%%%%%%%%%%%%%%%%%%%%%%%%%%%%%%%%%%%%%%
In conclusion, in this letter we derived the one-loop  
free energy in spinor QED  using the worldline 
formulation of quantum field theory. It turned out 
that the calculations follow very closely the zero temperature 
calculations. The only difference (at one loop) appears in 
applying the boundary conditions to the saddle point trajectory
that take into account the winding numbers around the compact
imaginary time direction. By applying the method, we were able
to established the general structure of the temperature dependent
part of the effective action in QED in an arbitrary external 
inhomogeneous magnetic field. Also, using the known result for 
the derivative expansion of the effective action at $T=0$, we 
derived the analogous expression at $T\neq 0$.

%%%%%%%%%%%%%%%%%%%%%%%%%%%%%%%%%%%%%%%%%%%%%%%%%%%%%%%%%%%%%%%%%%%%%%%%
We would like to thank V.~Gusynin for helpful comments on 
the manuscript. This work was supported in part by the 
U.S. Department of Energy Grant \#DE-FG02-84ER40153. 

%%%%%%%%%%%%%%%%%%%%%%%%%%%%%%%%%%%%%%%%%%%%%%%%%%%%%%%%%%%%%%%%%%%%%%%%

\end{document}